\documentclass[]{elsart}

\usepackage{natbib}

\usepackage{amssymb}
\usepackage{listings}
\usepackage{array}
\usepackage{longtable}
\usepackage{graphicx}
\begin{document}

\begin{frontmatter}



\title{Performance Tuning of $N$-Body Codes on Modern Microprocessors: \\
	I. Direct Integration with a Hermite Scheme on x86\_64 Architecture}


\author[Hongo]{Keigo Nitadori}
\author[Hongo]{Junichiro Makino}
\author[IAS]{Piet Hut}

\address[Hongo]{Department of Astronomy, University of Tokyo, 7-3-1 Hongo, Bunkyo-ku, Tokyo 113-0033, Japan}
\address[IAS]{Institute for Advanced Study, Princeton, NJ 08540, USA}

\begin{abstract}
The main performance bottleneck of gravitational $N$-body codes
is the force calculation between two particles.
We have succeeded in speeding up this pair-wise force calculation
by factors between two and ten, depending on the code and the
processor on which the code is run.  These speedups were obtained
by writing highly fine-tuned code for x86\_64
microprocessors.  Any existing $N$-body code, running on these chips,
can easily incorporate our assembly code programs.

In the current paper, we present an outline of our overall approach,
which we illustrate with one specific example: the use of a Hermite
scheme for a direct $N^2$ type integration on a single 2.0 GHz Athlon
64 processor, for which we obtain an effective performance of 4.05
Gflops, for double precision accuracy.  In subsequent papers, we will
discuss other variations, including the combinations of $N\log N$
codes, single precision implementations, and performance on other
microprocessors. 

\end{abstract}


\end{frontmatter}

\section{Introduction}

Some $N$-body simulations can be sped up in various ways, by using
faster algorithms such as tree codes \citep{BarnesHut1986} and/or
special purpose hardware such as the GRAPE family
\citep{Sugimotoetal1990}.  For some regimes, such as low $N$ values,
these speed-up methods are not very efficient, and it would be nice to
find other ways to improve the speed of such calculations.  It would
be even better if these alternative ways can be combined with 
other methods of speed-up.

We explore here a general approach based on speeding up the inner loop
of gravitational force calculations, namely the interactions between
one pair of particles.  Also when using tree codes this approach will
be useful, since in that case the calculation cost is still dominated
by force calculations. Even for GRAPE applications, this approach
will still be useful in many cases, since there are always some
calculations which are done more efficiently on the front end.

In particular, we consider the optimization of the inner force loop on
the x86-64 (or AMD64 or EM64T) architecture, the newest incarnation of
the architecture that originated with the Intel 8080 microprocessor.
Processors with an x86-64 instruction set are currently the most
widely used.  Athlon 64 and Opteron microprocessors from AMD, and many
recent models of Pentium 4 and Xeon microprocessors from Intel, support
this instruction set.

As will be shown in section \ref{sec:baseline}, a straightforward 
implementation of
the inner force loop using either Fortran or C, compiled with standard
compilers like GCC or icc for x86-64 processors, results in a performance
that is significantly lower than the theoretical peak value one can expect
from the hardware.  In the following, we discuss how we can improve the
performance of the force loop on processors with an x86-64 instruction set.

\subsection{The SSE2 Instruction Set}

Our approach is based on the use of new features added to the x86
microprocessors in the last eight years.  The 
first one is the SSE2 instruction set for double-precision
floating-point arithmetic.  Traditionally, the instruction set for x86
microprocessors has included the so-called x87 instruction set, which
was originally designed for the 8087 math coprocessor for the 8086
16-bit microprocessor.  This instruction set is stack-based, in the
sense that it does not have any explicit way to specify registers.
Instead, registers are indirectly accessed as a stack, where two
operands for an arithmetic operation are taken from the top of the
stack (popped) and the result is placed back at the top of the stack
(pushed).  Memory access also takes place through the top of the
stack.

This x87 instruction set had the advantage that the instructions are
simple and few in number, but for the last fifteen years the design of
a fast floating-point unit for this x87 instruction set has been a
major problem for all x86 microprocessors.  If one would really
design stack-based hardware, any pipelining would be
practically impossible.  In order to allow pipelining, current x86-based
microprocessors, from Intel as well as AMD, translate the stack-based x87
instructions to RISC-like, presumably standard three-address
register-to-register instructions in hardware at execution time.

This approach has given quite high performance, certainly much higher than
what would have been possible with the original stack-based implementation.
However, it was clear that pipelining and better use of hardware
registers would be much easier if one could use an instruction set
with explicit reference to the  registers.  In 2001, with
the introduction of Pentium 4 microprocessors, Intel added such a new
floating-point instruction set, which is called SSE2.  It is still not
a real three-address instruction set; it rather uses a two-address
form where the address of the source register and that of the destination
register are the same.  
Moreover, SSE2 still supports operations between the data in the main
memory and that in the register, as was the case with the IBM System/360.
Thus, it still has the look and feel of an instruction set from the 1960s.

\subsection{Minimizing Memory Access}

Even though operations between operands in memory and operands in
registers are supported, clearly execution would be much faster if all
operands could reside in registers.  However, with the original SSE2
instruction set it was difficult to eliminate memory access for
intermediate results, because there were only eight 
registers available for SSE2 instructions. For whatever reason, these
registers are called ``XMM'' registers in the manufacturer's document,
and we follow this convention. 
With the new x86\_64 instruction set, the number of these ``XMM''
registers was doubled from 8 to 16.  The implication for $N$-body
calculations was that it now became possible to minimize memory access
during the inner force loop.  A form of optimization using this approach
will be discussed in section \ref{sec:c-level}.

\subsection{Exploiting Two-Word Double-Precision Parallelism}

Another important feature of SSE2 (which is actually Streaming SIMD
extension 2) is that it is defined to operate on a pair of two 64-bit
floating point words, instead of a single floating-point word.  This
effectively means that the use of SSE2 instructions automatically
result in the execution of two floating-point operations in parallel.
While this feature cannot be easily used with any compiler-based
optimization, it is possible to gain considerable profit from this
feature through judicious hand coding.  We discuss the use of this
parallel nature of the SSE2 instruction set in section \ref{sec:sse2}.

\subsection{Exploiting Four-Word Single-Precision Parallelism}

SSE2 is not the only new floating-point instruction set that has been
made available for the x86 hardware.  As the name SSE2 already
suggests, there is an earlier SSE instruction set, which is similar to
SSE2 but works only on single-precision floating-point numbers.  As is
the case with SSE2, SSE also works on multiple data in parallel, but
instead of the two double-precision data of SSE2, SSE works on four
single-precision floating-point numbers simultaneously.  Thus, the peak
calculation speed of SSE is at least a factor of two higher than that of
SSE2.  For those force calculations where single precision gives us a
sufficient degree of accuracy, we can make use of SSE, gaining a
performance that is even higher than what would have been possible with 
SSE2, as we discuss in section \ref{sec:mixed}.

\subsection{Utilizing Built-In Inverse Square Root Instructions}

SSE was designed mainly to speed up coordinate transformation in
three-dimensional graphics.  As a result, it has a special instruction
for the very fast calculation of an approximate inverse square
root, which is intended as a good initial value for Newton-Raphson
iteration. This is exactly what we need for the calculation of
gravitational forces. 
We discuss the use of this approximate inverse square root 
for double-precision calculation, in section \ref{sec:sse2}.

\subsection{Organization}

This paper is organized as follows. In section \ref{sec:baseline}, we
give the standard C-language implementation of the force loop. We
consider the code fragment which calculates both the acceleration and
its first time derivative.  It is used with Hermite integration scheme
\citep{MakinoAarseth1992}.

We present the assembly language output and measured
performance, and describe possible room for improvements.  We call
these implementations the baseline implementation. 

In section \ref{sec:c-level}, we discuss the optimized C-language
implementation of
the force loop for the Hermite scheme.  The difference from the
baseline implementation is that the C-language code is hand-tuned so
that the load and store of intermediate results are eliminated from
the generated assembly-language output.  This version gives us the speed-up of 46\% compared
to the baseline method. 

In section 4, we discuss a more efficient use of SSE2 instruction,
where forces on two particles are calculated in parallel using the
SIMD nature of the SSE2 instruction set. Here, we use the intrinsic
data types defined in GCC, which allows us to use this SIMD nature
within the syntax of C-language. 
Also, we use the fast approximate
square root instruction and a Newton-Raphson iteration.
This implementation is 88\% faster than baseline.

In section 5, we discuss a mixed SSE-SSE2 implementation of the force
calculation for the Hermite scheme.  In many applications, full
double-precision accuracy is not necessary, except for the first
subtraction between positions and final accumulation of the forces.
"High-accuracy" GRAPE hardwares (GRAPE-2, 4 and 6) rely on this
mixed-accuracy calculation.  Thus, it is possible to perform most of
the force-calculation operations using SSE single-precision
instructions, thereby further speeding up the force calculation.  In
this way, we can speed-up the calculation by another 67 \% from the
SSE2 parallel implementation of section 4, achieving 219\% speedup
(3.19 times faster) from the baseline implementation.

\section{Baseline implementation}
\label{sec:baseline}

\subsection{Target functions}
\label{subsec:target}

The target functions that we want to calculate are:

\begin{eqnarray}
{\bf a}_i & = &
\sum_j {m_j {\bf r}_{ij} \over (r_{ij}^2 + \varepsilon^2)^{3/2}} \\
{\bf j}_i & = & {{\bf{\dot a}}_i} \ =\ 
\sum_j m_j \left[{{\bf v}_{ij} \over (r_{ij}^2 + \varepsilon^2)^{3/2}}
- {3({\bf v}_{ij}\cdot{\bf r}_{ij}){\bf r}_{ij} \over (r_{ij}^2
+ \varepsilon^2)^{5/2}} \right] \\
 \phi_i & = &
-\sum_j {m_j \over (r_{ij}^2 + \varepsilon^2)^{1/2}}
\end{eqnarray}
where ${\bf a}_i$ and $\phi_i$ are the gravitational acceleration and
the potential of particle $i$, the jerk ${\bf j}_i$ is the time
derivative of the acceleration, and ${\bf r}_i$, ${\bf v}_i$, and $m_i$
are the position, velocity and mass of particle $i$, ${\bf r}_{ij} =
{\bf r}_{j} - {\bf r}_{i}$ and ${\bf v}_{ij} = {\bf v}_{j} - {\bf
v}_{i}$.

The calculation of ${\bf a}$ and $\phi$ requires 9 multiplications, 10
addition/subtraction operations, 1 division and 1 square root
calculation.  Clearly, calculation of an inverse square root is more
expensive than addition/multiplication operations.  Therefore, if we
want to measure the speed of a force calculation loop in terms of the
number of floating-point operations per second, we need to introduce
some conversion factor for division and square root calculations.  In
this paper, we use 38 as the total number of floating point operations
for the calculation of ${\bf a}$
and $\phi$. 
This implies that we effectively
assign around 10 operations to each division and to each square root
calculation.  This particular convention was introduced by
\cite{WSB+}, and we follow it here since it seems to be a reasonable
representation of the actual computational costs of force calculations on
typical scalar processors.

In addition, it requires 11 multiplications and 11
addition/subtraction operations to calculate ${\bf j}$.  Thus, the
total number of floating-point operations per inner force loop for the
Hermite scheme can be given as 60.  This is the number that we will
use here.  Note that we have previously used numbers that were slightly
less, by about 5\%, using 57 instead of 60
\citep{G6SC2001}.

In the case of a simple leapfrog method or traditional "Aarseth"
scheme, we need to calculate only ${\bf a}_i$ and $\phi_i$.  For a Hermite
scheme we need to determine ${\bf j}_i$ as well.

\subsection{Baseline implementation and its performance}

The following code fragments contain what we regard as the baseline
implementation of the force calculations, where we use the word
`force' loosely, to indicate the calculations for accelerations,  jerks 
as well as the potential.  In other words, we consider `force
calculations' to comprise all the basic low-level dynamical
calculations governing the interactions between pairs of particles.

\renewcommand{\lstlistingname}{\small List}
\begin{lstlisting}[caption= {
	Baseline implementation calculates force on $i$th particle.
}, label=C:baseline]
#define DIM 3

typedef struct{
	double x[DIM];
	double v[DIM];
	double m;
}Predictor;
typedef Predictor * pPredictor;

typedef struct{
	double a[DIM];
	double j[DIM];
	double pot;
}AccJerk;
typedef AccJerk * pAccJerk;

void CalcAccJerk_Base (pPredictor pr, pAccJerk aj, 
		               double eps2, int i, int nbody){
    int j, k;
    double r[3], v[3], acc[3], jerk[3], pot;
    double r2, rv;
    double rinv,rinv2;
	double mrinv, mrinv3, rvrinv2;

    pot = 0.0;
    for(k=0; k<3; k++) acc[k] = jerk[k] = 0.0;

    for(j=0; j<nbody; j++){
        if(j == i) continue;
        for(k=0; k<3; k++){
            r[k] = pr[j].x[k] - pr[i].x[k];
            v[k] = pr[j].v[k] - pr[i].v[k];
        }

        r2 = eps2;
        for(k=0; k<3; k++) r2 += r[k] * r[k];
        rv = r[0] * v[0];
        for(k=1; k<3; k++) rv += r[k] * v[k];

        rinv2 = 1./r2;
        rinv = sqrt(rinv2);
        rvrinv2 = 3.0 * rv * rinv2;;
        mrinv = rinv * pr[j].m;
        mrinv3 = mrinv * rinv2;

        pot -= mrinv;

        for(k=0; k<3; k++){
            double temp = r[k] * mrinv3;
            acc [k] += temp;
            jerk[k] += v[k] * mrinv3 - temp * rvrinv2;
        }
    }

    for(k=0; k<3; k++){
        aj->a[k] = acc[k];
        aj->j[k] = jerk[k];
        aj->pot = pot;
    }
}
\end{lstlisting}
Line 29 in list \ref{C:baseline} is a branch to avoid self interaction.
We might remove this branch when we can use softening,
but in this case branch prediction of the microprocessor works ideally,
hence the performance changes little if we remove this. 

\begin{table}
\caption{ 
	Performance of the code shown in list \ref{C:baseline} when $N = 1024$,
	in cycles-per-interaction and Gflops, on Athlon 64 2.0GHz.
}
\label{table:baseline}

\begin{tabular}{ l >{}l r r}
\hline
\multicolumn{1}{c}{Compiler} &
\multicolumn{1}{c}{Options} &
\multicolumn{1}{c}{Cycles} &
\multicolumn{1}{c}{Gflops} 
\\ 
\hline
GCC 3.3.1  & -O3 -ffast-math -funroll-loops & 94.8 & 1.27 \\
GCC 3.3.1  & -O3 -ffast-math -funroll-loops -mfpmath=387 & 119 & 1.01 \\
GCC 4.0.1  & -O3 -ffast-math -funroll-loops & 100 & 1.20 \\
PGI 5.1  & -fastsse & 97.6 & 1.23 \\
PathScale 2.0 & -O3 -ffast-math & 95.0 & 1.26 \\
ICC 9.0 & -O3 & 95.8 & 1.25 \\
\hline
\end{tabular}
\end{table}

Table \ref{table:baseline} shows the performance of this code on an
AMD Athlon 64 3000+ (2.0GHz) processor with several different
compilers.  The first column gives the compiler used, the second the compiler
options, the third the clock cycles per pairwise force calculation, and
the fourth column gives the speed in Gflops.
All compilers generate SSE2 instructions instead of 
x87 instructions unless we explicitly set options to use x87. 
The performance is fairly good, but not ideal. In the following we will
investigate how we can improve the performance.

List \ref{asm:baseline} is the assembly language output of the Hermite force loop using
the GCC compiler:

\begin{lstlisting}[caption= 
{
	Assembly output of list \ref{C:baseline} using GCC 3.3.1 with option
	{\tt -S -O3 -ffast-math -funroll-loops}, commented by the
	authors for clarity.
}, label=asm:baseline]
	xorpd	%xmm9, %xmm9
	xorl	%r8d, %r8d
	subq	$16, %rsp
.LCFI0:
	movsd	%xmm0, %xmm10
	movl	%edx, %r11d
	cmpl	%ecx, %r8d
	movl	%ecx, %r10d
	movsd	%xmm9, -120(%rsp)
	movsd	%xmm9, -88(%rsp)
	movsd	%xmm9, -112(%rsp)
	movsd	%xmm9, -80(%rsp)
	movsd	%xmm9, -104(%rsp)
	movsd	%xmm9, -72(%rsp)
	jge	.L41
	movslq	%edx,%rdx
	movlpd	.LC4(%rip), %xmm11
	movlpd	.LC5(%rip), %xmm12
	leaq	0(,%rdx,8), %r9
	subq	%rdx, %r9
	.p2align 4,,7
.L60:
	cmpl	%r11d, %r8d
	je	.L9							# continue
	movslq	%r8d,%rcx
	leaq	0(,%rcx,8), %rdx
	subq	%rcx, %rdx				# %rdx = 7 * %rcx
	leaq	8(,%r9,8), %rcx
	movlpd	(%rdi,%rdx,8), %xmm6	# %xmm6 = pr[j].x[0]
	leaq	8(,%rdx,8), %rax
	subsd	(%rdi,%r9,8), %xmm6		# %xmm6 -= pr[i].x[0]
	movsd	%xmm6, -24(%rsp)		# *(%rsp-24) = %xmm6
	movsd	%xmm6, %xmm3
	movlpd	24(%rdi,%rdx,8), %xmm4
	subsd	24(%rdi,%r9,8), %xmm4	# %xmm4 = pr[j].v[0] - pr[i].v[0]
	mulsd	%xmm6, %xmm3			# %xmm3 = x*x
	addsd	%xmm10, %xmm3			# %xmm3 += eps2
	movsd	%xmm4, -56(%rsp)
	movlpd	(%rdi,%rax), %xmm7
	subsd	(%rdi,%rcx), %xmm7		# %xmm7 = pr[j].x[1] - pr[i].x[1]
	movsd	%xmm7, -16(%rsp)
	movsd	%xmm7, %xmm15
	movsd	%xmm7, %xmm2
	movlpd	24(%rdi,%rax), %xmm14
	leaq	16(,%rdx,8), %rax
	subsd	24(%rdi,%rcx), %xmm14	# %xmm14 = pr[j].v[1] - pr[i].v[1]
	mulsd	%xmm7, %xmm15
	leaq	16(,%r9,8), %rcx
	addsd	%xmm15, %xmm3			# %xmm3 += y*y
	movlpd	-88(%rsp), %xmm15		# load acc[0]
	movsd	%xmm14, -48(%rsp)
	mulsd	%xmm14, %xmm2
	movlpd	(%rdi,%rax), %xmm8
	subsd	(%rdi,%rcx), %xmm8		# %xmm8 = pr[j].x[2] - pr[i].x[2]
	movsd	%xmm8, %xmm5
	movsd	%xmm8, -8(%rsp)
	movsd	%xmm8, %xmm0
	movlpd	24(%rdi,%rax), %xmm13
	mulsd	%xmm8, %xmm5
	subsd	24(%rdi,%rcx), %xmm13	# %xmm13 = pr[j].v[2] - pr[i].v[2]
	addsd	%xmm5, %xmm3			# %xmm3 += z*z
	movsd	%xmm6, %xmm5
	mulsd	%xmm4, %xmm5			# %xmm5 = x*vx
	movsd	%xmm13, -40(%rsp)
	mulsd	%xmm13, %xmm0
	addsd	%xmm2, %xmm5			# %xmm5 += y*vy
	movsd	%xmm11, %xmm2			# %xmm2 = 1.0
	divsd	%xmm3, %xmm2			# %xmm2 = 1.0/r2
	movlpd	-80(%rsp), %xmm3
	addsd	%xmm0, %xmm5			# %xmm5 += z*vz
	sqrtsd	%xmm2, %xmm1			# %xmm1 = 1.0/r
	mulsd	%xmm2, %xmm5			# %xmm5 = rv/r2
	mulsd	48(%rdi,%rdx,8), %xmm1	# %xmm1 *= pr[j].m
	mulsd	%xmm12, %xmm5			# %xmm5 = 3rv/r2
	mulsd	%xmm1, %xmm2			# %xmm2 = m/r3
	subsd	%xmm1, %xmm9			# pot -= m/r
	movlpd	-72(%rsp), %xmm1
	mulsd	%xmm2, %xmm6
	mulsd	%xmm2, %xmm7
	mulsd	%xmm2, %xmm8
	mulsd	%xmm2, %xmm4
	addsd	%xmm6, %xmm15			# acc[0] += x[0] * m/r3
	mulsd	%xmm2, %xmm14
	addsd	%xmm7, %xmm3
	mulsd	%xmm5, %xmm6
	addsd	%xmm8, %xmm1
	mulsd	%xmm5, %xmm7
	mulsd	%xmm2, %xmm13
	mulsd	%xmm5, %xmm8
	movsd	%xmm15, -88(%rsp)		# store acc[0]
	subsd	%xmm6, %xmm4
	movsd	%xmm3, -80(%rsp)		# store acc[1]
	subsd	%xmm7, %xmm14
	addsd	-120(%rsp), %xmm4
	movsd	%xmm1, -72(%rsp)		# store acc[2]
	addsd	-112(%rsp), %xmm14
	subsd	%xmm8, %xmm13
	addsd	-104(%rsp), %xmm13
	movsd	%xmm4, -120(%rsp)		# store jerk[0]
	movsd	%xmm14, -112(%rsp)		# store jerk[1]
	movsd	%xmm13, -104(%rsp)		# store jerk[2]
.L9:
	incl	%r8d
	cmpl	%r10d, %r8d
	jl	.L60						# if(++j < nbody) goto .L60:
.L41:
	movq	-88(%rsp), %r11
	movq	-120(%rsp), %r10
	movsd	%xmm9, 48(%rsi)
	movq	-80(%rsp), %r9
	movq	-112(%rsp), %r8
	movq	-72(%rsp), %rdi
	movq	-104(%rsp), %rdx
	movq	%r11, (%rsi)
	movq	%r10, 24(%rsi)
	movq	%r9, 8(%rsi)
	movq	%r8, 32(%rsi)
	movq	%rdi, 16(%rsi)
	movq	%rdx, 40(%rsi)
	addq	$16, %rsp
	ret
\end{lstlisting}

One can see that there are quite a few unnecessary load/store
instructions. 
For example, the store instructions in lines 32, 38, 41, 51, 56 and 64
of list \ref{asm:baseline} serve no purpose.
Also Arrays {\tt acc[3]} and {\tt jerk[3]} could be placed in registers
like variable {\tt pot}, but they are placed in memory instead.

\section{C-level optimization}
\label{sec:c-level}

\subsection{Code modification}

As we have seen in the previous section, the assembly language output
shows a significant amount of unnecessary load/store instructions.  In
principle, if the compilers were clever enough they would be able to
eliminate these unnecessary operations.  In practice, we need to guide
the compilers so that they do not generate unnecessary code.

We have achieved significant speedup using the following two guiding
principles:

\begin{enumerate}

\item Eliminate assignment to any array element in the force loop.

\item Reuse variables as much as possible in order to minimize the
number of registers used.

\end{enumerate}

Apparently, present-day compilers are not clever enough to eliminate
load/store operations, if elements of arrays are used as left-hand-side
values.  Therefore, we hand-unroll all loops of length three and use
scalar variables instead of arrays for three-dimensional vectors.

In well-written programs, variables which contain the values for
different physical quantities should have different names.  However,
the use of many different names prevents optimization, since it
results in a number of variables too large to be fitted into the
register set of SSE2 instructions (there are 16 "XMM" registers for
SSE2 instructions).  Therefore, we explicitly reuse variables such as
{\tt x, y, z} so that these can hold both the position difference
components as well as the pairwise force components, at different
times in the computation.

The resulted C-language code is the following:

\begin{lstlisting}[caption= 
{
	An optimized force loop.
}, label=C:scalar]
typedef struct{
	double x[3];
	double v[3];
	double m;
	double pad;
}Predictor;
typedef Predictor * pPredictor;

void CalcAccJerk_Fast (pPredictor pr, pAccJerk aj, 
                       double eps2, int i, int nj){
    int j;
    double x,y,z,vx,vy,vz;
	double ax,ay,az,jx,jy,jz,pot;
    double r2,rv;
    double rinv,rinv2,rinv3;
    pPredictor jpr = pr;
    pPredictor ipr = pr + i;

    pot=ax=ay=az=jx=jy=jz=0.0;

    for(j=0;j<nj;j++,jpr++){
        __builtin_prefetch(jpr+2, 0, 3);
        if(j == i) continue;
        x = jpr->x[0] - ipr->x[0];
        y = jpr->x[1] - ipr->x[1];
        z = jpr->x[2] - ipr->x[2];

        vx= jpr->v[0] - ipr->v[0];
        vy= jpr->v[1] - ipr->v[1];
        vz= jpr->v[2] - ipr->v[2];

        r2 = eps2 + x*x + y*y + z*z;

        rv = vx*x + vy*y + vz*z;

        rinv2 = 1./r2;
        rinv = sqrt(rinv2);
        rv *= rinv2*3.;
        rinv *= jpr->m;
        rinv3 = rinv * rinv2;

        pot -= rinv;

        x *= rinv3; ax += x;
        y *= rinv3; ay += y;
        z *= rinv3; az += z;

        vx *= rinv3; jx += vx;
        vy *= rinv3; jy += vy;
        vz *= rinv3; jz += vz;

        x  *= rv; jx -= x;
        y  *= rv; jy -= y;
        z  *= rv; jz -= z;
    }

    aj->a[0] = ax;
    aj->a[1] = ay;
    aj->a[2] = az;
    aj->j[0] = jx;
    aj->j[1] = jy;
    aj->j[2] = jz;
    aj->pot = pot;
}
\end{lstlisting}

This code is compiled into the following assembly code using the GCC 3.3.1
compiler (flag {\tt -O3 -ffast-math}):

\begin{lstlisting}[caption={
	Assembly output of list \ref{C:scalar}.
}, label=asm:scalar]
CalcAccJerk_Fast:
.LFB3:
	movslq	%edx,%r8
	xorpd	%xmm9, %xmm9
	salq	$6, %r8
	movsd	%xmm0, -8(%rsp)
	leaq	(%r8,%rdi), %rax
	movsd	%xmm9, %xmm11
	movsd	%xmm9, %xmm10
	xorl	%r8d, %r8d
	movsd	%xmm9, %xmm14
	movsd	%xmm9, %xmm13
	cmpl	%ecx, %r8d
	movsd	%xmm9, %xmm12
	movsd	%xmm9, %xmm15
	jge	.L9
	.p2align 4,,7
.L11:
	cmpl	%edx, %r8d
	prefetcht0	128(%rdi)
	je	.L4
	movlpd	(%rdi), %xmm3
	movlpd	8(%rdi), %xmm4
	subsd	(%rax), %xmm3
	movlpd	16(%rdi), %xmm5
	subsd	8(%rax), %xmm4
	movlpd	24(%rdi), %xmm6
	subsd	16(%rax), %xmm5
	movlpd	32(%rdi), %xmm7
	subsd	24(%rax), %xmm6
	movlpd	40(%rdi), %xmm8
	subsd	32(%rax), %xmm7
	subsd	40(%rax), %xmm8
	movsd	%xmm3, %xmm0
	movsd	%xmm4, %xmm2
	mulsd	%xmm3, %xmm0
	addsd	-8(%rsp), %xmm0
	mulsd	%xmm4, %xmm2
	movsd	%xmm7, %xmm1
	mulsd	%xmm4, %xmm1
	addsd	%xmm2, %xmm0
	movsd	%xmm5, %xmm2
	mulsd	%xmm5, %xmm2
	addsd	%xmm2, %xmm0
	movsd	%xmm6, %xmm2
	mulsd	%xmm3, %xmm2
	addsd	%xmm1, %xmm2
	movsd	%xmm8, %xmm1
	mulsd	%xmm5, %xmm1
	addsd	%xmm1, %xmm2
	movlpd	.LC4(%rip), %xmm1
	divsd	%xmm0, %xmm1
	sqrtsd	%xmm1, %xmm0
	mulsd	%xmm1, %xmm2
	mulsd	48(%rdi), %xmm0
	mulsd	.LC5(%rip), %xmm2
	mulsd	%xmm0, %xmm1
	subsd	%xmm0, %xmm15
	mulsd	%xmm1, %xmm3
	mulsd	%xmm1, %xmm4
	mulsd	%xmm1, %xmm5
	mulsd	%xmm1, %xmm6
	mulsd	%xmm1, %xmm7
	addsd	%xmm3, %xmm12
	mulsd	%xmm1, %xmm8
	addsd	%xmm4, %xmm13
	addsd	%xmm5, %xmm14
	mulsd	%xmm2, %xmm3
	addsd	%xmm6, %xmm10
	mulsd	%xmm2, %xmm4
	addsd	%xmm7, %xmm11
	mulsd	%xmm2, %xmm5
	addsd	%xmm8, %xmm9
	subsd	%xmm3, %xmm10
	subsd	%xmm4, %xmm11
	subsd	%xmm5, %xmm9
.L4:
	incl	%r8d
	addq	$64, %rdi
	cmpl	%ecx, %r8d
	jl	.L11
.L9:
	movsd	%xmm12, (%rsi)
	movsd	%xmm13, 8(%rsi)
	movsd	%xmm14, 16(%rsi)
	movsd	%xmm10, 24(%rsi)
	movsd	%xmm11, 32(%rsi)
	movsd	%xmm9, 40(%rsi)
	movsd	%xmm15, 48(%rsi)
	ret
\end{lstlisting}

\begin{figure}[htpd]
\begin{small}
\begin{center}
\begin{tabular}{| l | >{\tt}l |}
\hline
\%xmm0 & rinv, r2 \\ \hline
\%xmm1 & rinv2, rinv3, y*vy, z*vz\\ \hline
\%xmm2 & rv, y*y, z*z \\ \hline
\%xmm3 & x \\ \hline
\%xmm4 & y \\ \hline
\%xmm5 & z \\ \hline
\%xmm6 & vx \\ \hline
\%xmm7 & vy \\ \hline
\%xmm8 & vz \\ \hline
\%xmm9 & jz \\ \hline
\%xmm10 & jx \\ \hline
\%xmm11 & jy \\ \hline
\%xmm12 & ax \\ \hline
\%xmm13 & ay \\ \hline
\%xmm14 & az \\ \hline
\%xmm15 & pot \\ \hline
\end{tabular}
\end{center}
\end{small}
\caption{The use of XMM registers in list \ref{asm:scalar}.}
\label{fig:regmap}
\end{figure}

\begin{table}[htpd]
\caption{
	Performance of list \ref{C:scalar} when $N = 1024$, 
	in cycles-per-interaction and Gflops, on Athlon 64 3000+ 2.0GHz. 
}
\begin{tabular}{l >{\tt}l r r}
\hline
\multicolumn{1}{c}{Compiler} &
\multicolumn{1}{c}{Options} &
\multicolumn{1}{c}{Cycles} &
\multicolumn{1}{c}{Gflops} 
\\ 
\hline
GCC 3.3.1  & -O3 -ffast-math & 64.8 & 1.85 \\
GCC 4.0.1  & -O3 -ffast-math & 64.7 & 1.85 \\
ICC 9.0  & -O3 & 79.4 & 1.51 \\
\hline
\end{tabular}
\label{table:scalar}
\end{table}

We can see that now all load/store instructions of the intermediate
results are eliminated.  Figure \ref{fig:regmap} shows the use of
registers and table \ref{table:scalar} presents the performance data for
this code. This version is 46\% faster than the code in section
\ref{sec:baseline}.  Very roughly, this speed-up is consistent with
the reduction in the number of assembly language instructions (from
82 to 62), but is somewhat larger.  This is partly because the
instructions which take memory arguments have a larger latency than
register-only instructions.

\subsection{Prefetch insertion and alignment}

Line 22 of list \ref{C:scalar} shows a built-in function for GCC,
which is compiled into a prefetch instruction.  In this case, the
prefetch instruction loads the data which will be needed two
iterations after the current one.  The second and third parameters are
read/write flags for which zero indicates preparing for a read and
the degree of temporal locality takes value from 1 to 3.

Line 6 of list \ref{C:scalar} shows a pad to make
the size of the predictor structure to be exactly 64-byte, which is the
cache-line size of Athlon 64 processor.
To make the predictor structure align at 64-byte boundary,
we  use {\tt memalign()} or {\tt posix\_memalign()} 
instead of {\tt malloc()}.

\begin{figure}[htbp]
\centering
\includegraphics[height=4in, angle=270]{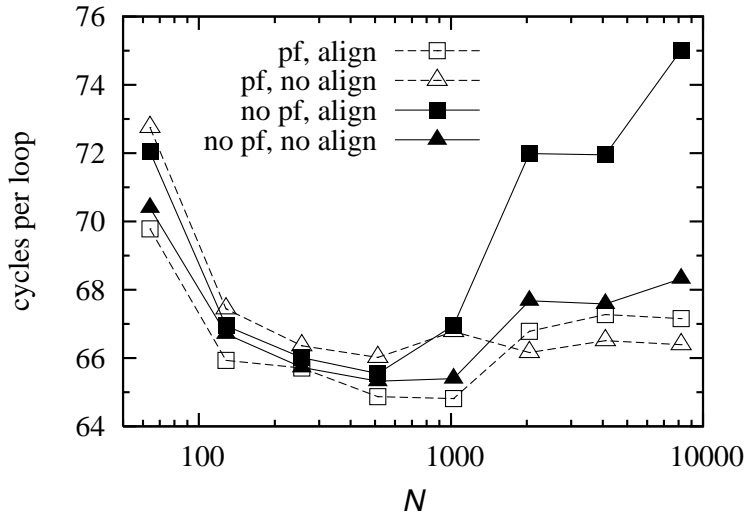}
\caption{
	Clock cycles per force calculation loop as a function of $N$
	on Athlon 64 3000+ 2.0 GHz. Open squares and triangles
show the results of using code with prefetch and with and without 64-byte
alignment. Filled squares and triangles show the
result of using code without prefetch and with and without 64-byte
alignment.  
}
\label{fig:prefetch}
\end{figure}

Figure \ref{fig:prefetch} shows the performance of the force loop as
the function of the loop length $N$, with and without this prefetch
instruction and 64-byte alignment. 
The fastest case depends on $N$. For the region $N \leq 1024$, 
in which data fits into 64 kB of L1 data cache, the code with
prefetch and 64-byte alignment is the best.
For larger $N$, the code with prefetch and without alignment
is the best.

\section{Assembly-level optimization}
\label{sec:sse2}

In this section, we give two extensions
specialized for x86/x86\_64 architecture for the force loop.
The first is using SSE2 vector (SIMD) mode instead of SSE2 scalar (SISD) mode.
The second is replacing one division and one square root,
by a special instruction for fast approximate inverse square root in SSE
and Newton-Raphson iteration.

\subsection{SSE2 vector mode}

In the previous section we improved the performance of the C-language
implementation of the force loop, essentially by hand-optimizing the C
code so that the generated assembly code becomes optimal. However, in
this way, we did not use the full capability of SSE2.  While SSE2
instructions can process two double-precision numbers in parallel, the
force loop discussed in the previous section uses only one of these
two words.  Clearly, we did not yet use the "SIMD" nature of the
instruction set.

Whether or not we can gain by using this SIMD nature depends on the
particular code, and also on the particular processor. On Intel P4
architecture the SIMD mode can offer up to a factor of two speedup, while
on AMD Athlon 64 or Intel Pentium M, the speed increase can be small (or
even negative).

There are many ways to use this SIMD feature.  Since there are
similarities with the vector instructions of some old vector
processors, in particular the Cyber 205, one could make use of an
automatic vectorizing compiler (such as icc version 6.0 and later).
However, just as was the case with the old vectorizing compilers, the
vectorizing capability of modern compilers are still very limited, and
it is hard to rewrite the force loop so that the compiler can make use
of the SIMD capability.

In fact, part of the reason why vectorization is difficult is that
the load/store capability of SSE2 instructions is rather limited:
it can work efficiently only on a pair of two double-precision words in
consecutive and 16-byte aligned addresses. 
This means that the basic loop structure
cannot work, and one needs to copy the data of two particles into
some special data structure, in order to let the compiler generate
the appropriate SIMD instructions.

Here we have adopted a low-level approach, in which we make use of the
special data type of pairs of double precision words defined in GCC.
Basically, this data type, which we call {\tt v2df}, corresponds to what is
in one XMM register (a pair of double-precision words).  We can perform
the usual arithmetic operations and even function calls for this data
type.  Here is the code which makes use of this {\tt v2df} data type:

\begin{lstlisting}[caption={Defining SSE/SSE2 data type.}]
typedef double v2df __attribute__ ((mode(V2DF)));
typedef float  v4sf __attribute__ ((mode(V4SF)));
\end{lstlisting}
The {\tt v4sf} data type packs four single-precision words
for SSE which we will use later.

The basic idea here is to calculate the forces from one particle on two
different particles in parallel.  One could try to calculate the forces from,
rather than on, two different particles on one particle in parallel,
but that would result in a more complicated program since then we will
need to add up the two partial forces in the end.  Also, from the
point of view of memory bandwidth, our approach is more efficient,
since we need to load only one particle per iteration.  Note that this
is the same approach as what is called $i$-parallelism in the various
versions of GRAPE hardware, where parallel pipelines calculate the
forces on different particles from the same set of particles.

In this code, we use macros for gathering load and scattering store using
instructions for loading/storing the higher/lower half of an XMM register:

\begin{lstlisting}[caption={Gathering load and scattering store for SSE2.}
,label={list:gather}]
#define V2DF_GATHER(reg, ptr0, ptr1) \
	reg = __builtin_ia32_loadlpd(reg, (void *)(ptr0)); \
	reg = __builtin_ia32_loadhpd(reg, (void *)(ptr1));

#define V2DF_SCATTER(reg, ptr0, ptr1) \
	__builtin_ia32_storelpd((void *)(ptr0), reg); \
	__builtin_ia32_storehpd((void *)(ptr1), reg);
\end{lstlisting}

We should be careful about the fact that these built-in functions are
undocumented feature of GCC.  The GCC document describes built-in functions
for most SSE/SSE3 instructions but not for any of SSE2 instruction.  We can
call most SSE2 instruction through the prefix {\tt
\_\_builtin\_ia32\_}.  Note that, for whatever reason, instructions
such as "movxxx" from/to memory are changed to "loadxxx"/"storexxx".
But this does not mean that GCC always generates correct assembly
output.  For example, GCC generates wrong code for the first macro in
list \ref{list:gather} when {\tt reg} is not a register variable.
This can be considered either a bug or a feature, and it shows the
risk of using these undocumented aspects of GCC.

We also use macros for numerical values since we cannot use
numerical values like "3.0" for vector operations:
\begin{lstlisting}[caption={
	Macros for numerical values. 
}]
#define V2DF_OP_SCALAR(reg, op, imm) { \
	v2df v2df_temp = {imm, imm}; \
	reg op v2df_temp; \
}
#define V2DF_OP_VECTOR(reg, op, imm0, imm1) { \
	v2df v2df_temp = {imm0, imm1}; \
	reg op v2df_temp; \
}
#define V4SF_OP_SCALAR(reg, op, imm) { \
	v4sf v4sf_temp = {imm, imm, imm, imm}; \
	reg op v4sf_temp; \
}
\end{lstlisting}

Note that we need to copy the data of a single particle into both
high and low
words of
an XMM register. The new SSE3 extension supports this "broadcasting",
while the original SSE2 set did not.  Therefore, we use the following macro:

\begin{lstlisting}[caption={Broadcast loading for double precision.}]
#ifdef __SSE3__
#define LOADDDUP(reg, ptr) \
	reg = __builtin_ia32_loadddup(ptr);
#else
#define LOADDDUP(reg, ptr) \
	reg = __builtin_ia32_loadlpd(reg, (void *)(ptr)); \
	reg = __builtin_ia32_loadhpd(reg, (void *)(ptr));
\end{lstlisting}

We show the vectorized force loop using these data types and
macros in list \ref{C:vector}.
\begin{lstlisting}[caption={
	Vectorized force loop using SSE2 vector mode.
}, label = {C:vector}]
typedef struct{
	double x[3];
	double v[3];
	double m[2];
}Predictor;
typedef Predictor * pPredictor;

void CalcAccJerk_Vector (pPredictor pr, pAccJerk AccJerkOut,
		                 double eps2, int index[], int nj){
    int j, k;
    v2df x,y,z,vx,vy,vz,ax,ay,az,jx,jy,jz,pot;
    v2df r2,rv;
    v2df rinv1,rinv2,rinv3;
    static v2df zero = {0.0, 0.0};
    pPredictor prj = pr;
    int i0 = index[0];
    int i1 = index[1];
    v2df xi[7];
    v2df *vi    = xi+3;
    v2df *eps2p = xi+6;

    pot=ax=ay=az=jx=jy=jz = zero;

    for(k=0;k<3;k++){
        V2DF_OP_VECTOR(xi[k], =, pr[i0].x[k], pr[i1].x[k]);
        V2DF_OP_VECTOR(vi[k], =, pr[i0].v[k], pr[i1].v[k]);
    }
    V2DF_OP_SCALAR(*eps2p, =, eps2);
    for(j=0;j<nj;j++,prj++){
        LOADDDUP(x, prj->x);
        x -= xi[0];
        LOADDDUP(y, prj->x+1);
        y -= xi[1];
        LOADDDUP(z, prj->x+2);
        z -= xi[2];
        LOADDDUP(vx, prj->v);
        vx -= vi[0];
        LOADDDUP(vy, prj->v+1);
        vy -= vi[1];
        LOADDDUP(vz, prj->v+2);
        vz -= vi[2];

        r2 = x*x + y*y + z*z + *eps2p;
        rv = vx*x + vy*y + vz*z;
        __builtin_prefetch(prj+2, 0, 3);

        V2DF_OP_SCALAR(rinv2, =, 1.0);
        rinv2 /= r2;
        rinv1 = __builtin_ia32_sqrtpd(rinv2);

        rv *= rinv2;
        V2DF_OP_SCALAR(rv, *=, 3.0);

        rinv1 *= *(v2df *)(prj->m);
        pot -= rinv1;
        rinv3 = rinv1 * rinv2;

        x *= rinv3; ax += x;
        y *= rinv3; ay += y;
        z *= rinv3; az += z;

        vx *= rinv3; jx += vx;
        vy *= rinv3; jy += vy;
        vz *= rinv3; jz += vz;

        x  *= rv; jx -= x;
        y  *= rv; jy -= y;
        z  *= rv; jz -= z;
    }

    {   /* correct self interaction */
        v2df mass = {pr[i0].m[0], pr[i1].m[0]};
        V2DF_OP_SCALAR(mass, *=, 1.0/sqrt(eps2));
        pot += mass;
    }
    V2DF_SCATTER(ax, AccJerkOut[0].a,   AccJerkOut[1].a);
    V2DF_SCATTER(ay, AccJerkOut[0].a+1, AccJerkOut[1].a+1);
    V2DF_SCATTER(az, AccJerkOut[0].a+2, AccJerkOut[1].a+2);
    V2DF_SCATTER(jx, AccJerkOut[0].j,   AccJerkOut[1].j);
    V2DF_SCATTER(jy, AccJerkOut[0].j+1, AccJerkOut[1].j+1);
    V2DF_SCATTER(jz, AccJerkOut[0].j+2, AccJerkOut[1].j+2);
    V2DF_SCATTER(pot, &AccJerkOut[0].pot, &AccJerkOut[1].pot);
}
\end{lstlisting}
The predictor structure has changed to store the same mass value in
two places (line 4), in order to save an instruction (line 54).

\subsection{Fast approximate inverse square root}

The most expensive part of the force calculation, for recent
microprocessors, is the calculation of an inverse square root
which requires one division and one square root caluclations.
Several
attempts to speed this up using table lookup, polynomial approximation
and Newton-Raphson iteration have been reported (e.g. Karp 1992). The
main difficulty in these approaches is how to get a good approximation
for the starting value of a Newton-Raphson iteration quickly.

Here, we use RSQRTSS/PS instruction in the SSE instruction set, 
which provide approximate
values for the inverse square root for scalar/vector single-precision
floating-point numbers, to an accuracy of about 12 bits.  
With one Newton-Raphson
iteration, we can obtain 24-bit accuracy, which is sufficient for most
``high-accuracy'' calculations. If higher accuracy is really
necessary, we could apply a second iteration.
The Newton-Raphson iteration formula is expressed as:
\begin{eqnarray}
\label{eqn:NR}
x_1 = -\frac{1}{2} x_0 (ax_0^2 - 3).
\end{eqnarray}
Here, $x_0$ is an initial guess for $1/\sqrt{a}$. 

We show the implementation for a scalar version using
inline assembly code with GCC extensions:

\begin{lstlisting}[caption={Calling {\tt rsqrtss} through inline
assembly code.}]
	double r2, rinv;
	float ftmp;

	ftmp = (float)r2;
	asm("rsqrtss %1, %0":"=x"(ftmp):"x"(ftmp));
	rinv = (double)ftmp;
	rinv *= r2*rinv*rinv - 3.0;
\end{lstlisting}
and for a vector version using built-in functions:
\begin{lstlisting}[caption={Calling {\tt rsqrtps} using built-in
functions of GCC.}]
	v2df r2, rinv;
	v4sf ftmp;

	ftmp = __builtin_ia32_cvtpd2ps(r2);
	ftmp = __builtin_ia32_rsqrtps(ftmp);
	rinv = __builtin_ia32_cvtps2pd(ftmp);
	r2 *= rinv;
	r2 *= rinv;
	V2DF_OP_SCALAR(r2, -= , 3.0);
	rinv *= r2;
\end{lstlisting}
Note that we skip here the multiplication by $-1/2$ in equation
(\ref{eqn:NR}). This can be done after the total force is
obtained. 

To use RSQRTSS/PS instructions, we first convert $r^2$ into
single precision, then apply these instructions, and finally convert the
result back to double precision.

\begin{figure}[htbp]
\centering
\includegraphics[height=5in, angle=270]{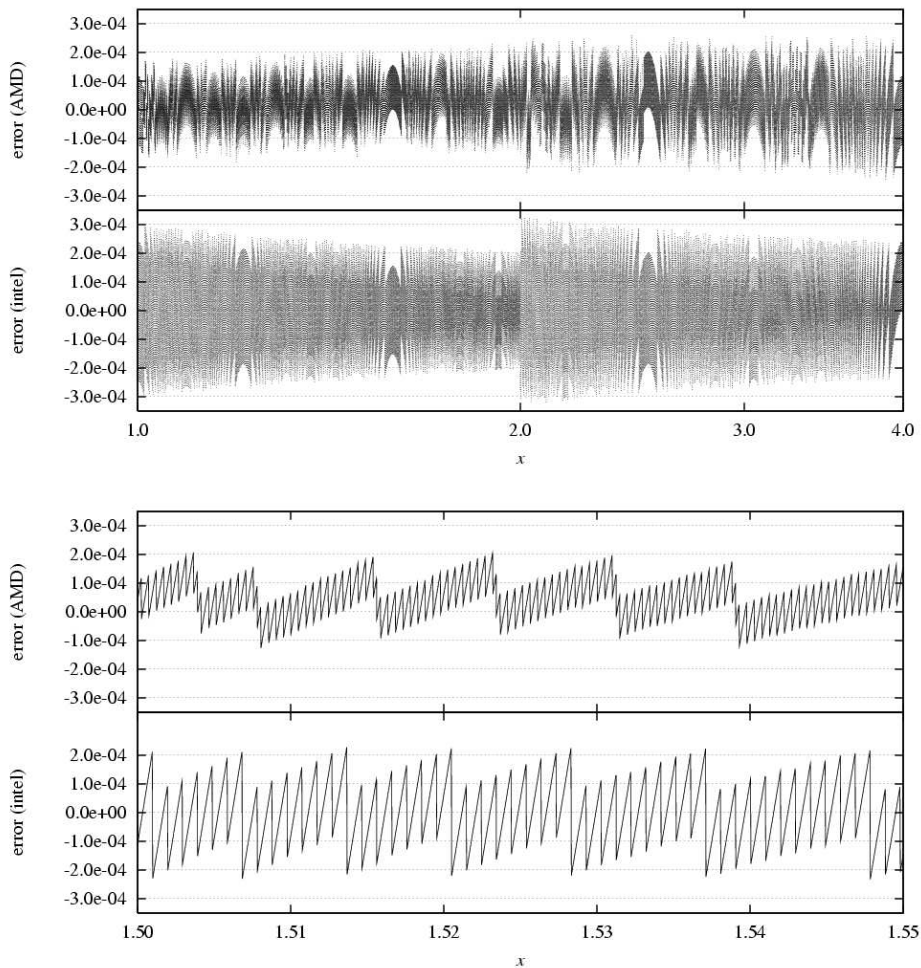}
\caption{
	Relative error of the return value of RSQRTSS/PS instructions on
	AMD Athlon 64  and Intel
	Pentium 4  for $1 \le x \le$ 16 (upper) and $1.5 \le x \le 1.55$ (lower).
	The errors is periodic for powers of 4.
}
\label{fig:rsqrt}
\end{figure}

Note that the actual value returned by this RSQRTSS/PS instruction is
implementation dependent.  In particular, the AMD Athlon 64 processors
and the Intel Pentium 4 processors return different values. 
Figure \ref{fig:rsqrt}
shows the errors of the return values as a function of the
input values.  
The AMD implementation has smaller average errors, but at the
same time they show a relatively large systematic bias.
Even after one Newton-Raphson iteration in double precision,
results from both implementations show relatively large biases.
Table \ref{table:error} presents the
root mean square error, max error, and bias (mean error)
of the approximate value $1/\sqrt{x}$, on a Intel Pentium 4 and an
AMD Athlon 64 before/after one Newton-Raphson iteration.
Here the error is measured as $1/2(x[\textrm{rsqrt}(x)]^2-1)$, and the
weight is $1/\textrm{frac}(x)$,
where $\textrm{frac}(x)$ is the normalized fraction of the floating-point number $x$.
We can correct these biases, at least statistically, by multiplying
the resulted force by constants which depend on the processor type used.

\begin{table}
\caption{ Errors of approximate inverse square root.
}
\newcommand{\expten}[2]{$#1\times 10^{#2}$}
\begin{tabular}{l r r r}
\hline
& 
\multicolumn{1}{c}{RMSE} & 
\multicolumn{1}{c}{MAX error} & 
\multicolumn{1}{c}{Bias} \\ 
\hline
AMD, before N-R &
$7.96\times 10^{-5}$ & $2.59\times 10^{-4}$ & $2.21\times 10^{-5}$ \\
AMD, after  N-R &
\expten{1.57}{-8} & \expten{1.00}{-7} & \expten{-9.51}{-9} \\ 
Intel, before N-R &
\expten{1.16}{-4} & \expten{3.26}{-4} & \expten{-8.37}{-8} \\
Intel, after  N-R &
\expten{3.05}{-8} & \expten{1.60}{-7} & \expten{-2.01}{-8} \\
IEEE 754 Single&
$3.58\times 10^{-8}$ & $8.94\times 10^{-8}$ & $1.52\times 10^{-11}$ \\
\hline
\end{tabular}
\label{table:error}
\end{table}

\subsection{Performance}
Table \ref{table:ScalarVector} shows the performance of four
different types of force loops, using SSE2 scalar/vector mode and
with/without fast inverse square root.

\begin{table}
\caption{
	Performance of SSE2 scalar mode and vector mode.
}
\label{table:ScalarVector}
\begin{tabular}{l r r r r}

\hline
SSE2 mode & \multicolumn{2}{c}{scalar} & \multicolumn{2}{c}{vector} \\
sqrt operation	 & normal & fast & normal & fast \\
\hline
Cycles per interaction & 64.8 cycle & 70.5 cycle & 69.0 cycle & 50.0 cycle \\
Calculation speed  & 1.85 Gflops & 1.70 Gflops & 1.73 Gflops & 2.40 Gflops \\ 
\hline
\end{tabular}
\end{table}

\section{Mixed-precision force loop}
\label{sec:mixed}

As we have summarized in the introduction, SSE2 is the SIMD
instruction set for pairs of double-precision words.  There are also
SSE instructions that work on quadruples of single-precision
words.  Thus, using the SSE instruction set, we can in principle double the
performance. 
If we perform the initial subtraction between positions and final
accumulation of acceleration and potential
in double-precision, we can use single-precision SSE
functions for all other calculations,
including subtraction between velocities and accumulation of jerk,
and still maintain a pretty high accuracy.  
The main complexity here is the question of how to make use
of four elements of SSE data type, in parallel.  The simplest approach
is to calculate the forces on four particles, but in that case we need
too many variables, which do not all fit into the registers. 
We need two XMM registers for each element of force in this way.
Instead, we have tried to achieve the maximum speed by
calculating the forces from two particles on two particles (making a
total of four force calculations) in parallel in SSE.

In the following, we present a detailed description of the
implementation of this mixed-precision force loop using SSE/SSE2.  
We use the term $i$ particles for particles which feel the
gravitational force, and $j$ particles for particles which exert the
force. 

\subsection{The data structure for $j$ particles}

The data structure for the $j$ particles should contain two particles.
The code in list \ref{list:mixedjp} achieves this goal by using the data
types {\tt v2df} and {\tt v4sf}. Note that for velocities we use the
single-precision {\tt v4sf} data type, since we do not need
double-precision accuracy for the velocity.  We store the data of the two
particles (with indices $j$ and $j+1$) as $(j, j+1)$ for position,
and s $(j, j, j+1, j+1)$ for velocity and mass. 
\begin{lstlisting}[caption={The $j$ particle structure.}, label={list:mixedjp}]
typedef struct{
	v2df x[3];
	v4sf v[3];
	v4sf m;
	v4sf pad;
}Jppack; /* 128-byte */
typedef Jppack * pJppack;
\end{lstlisting}
The variable {\tt pad} is used to make the size of the structure an
exact multiple of 64 bytes.

\subsection{The data structure for $i$ particles}

We use the following local variables to store two $i$ particles (with
indices $i_0$ and $i_1$).
\begin{lstlisting}[caption={Two $i$ particles packed into local variables.}]
v2df xi0[3] = {{x[i0][0], x[i0][0]}, {...} , {...}};
v2df xi1[3] = {{x[i1][0], x[i1][0]}, {...} , {...}};
v4sf vi[3]  = {{v[i0][0], v[i1][0], v[i0][0], v[i1][0]}, {...}, {...}};
\end{lstlisting}
Note that {\tt xi0} keeps the position of particle $i_0$ in a
duplicated way (two 64-bit words of v2df data type store the same
data). Similarly, {\tt xi1} keeps the data for $i_1$. In this way, by
subtracting {\tt x} of the $j$ particle variable from {\tt x0} or {\tt x1}, we can
calculate the displacement of two $j$ particles from one $i$
particle.  The results are then converted to single-precision format using a 
{\tt cvtpd2ps} instruction. (See list \ref{list:subpd}).
After {\tt unpcklps} (an instruction to pack two words into one
words,  not to unpack) is issued,
the contents of {\tt x} become 
$\{ x_j - x_{i0},\  x_j - x_{i1},\ x_{j+1} - x_{i0},\ x_{j+1} - x_{i1}\}$.

For the velocity, we store the data in the order of $(i_0, i_1,
i_0, i_1)$, so that a single subtraction operation provides the four
displacement vectors of two $j$ particles from two $i$ particles.

\begin{lstlisting}[caption={Subtraction between positions.}, label={list:subpd}]
pJppack pj;
v2df xi0[3], xi1[3];

for(j=0;j<n;j+=2, pj++){
	v4sf x, y, z, ftmp;
	x     = __builtin_ia32_cvtpd2ps(pj->x[0] - xi0[0]); 
	ftmp  = __builtin_ia32_cvtpd2ps(pj->x[0] - xi1[0]);
	x     = __builtin_ia32_unpcklps(x, ftmp); 
\end{lstlisting}

\subsection{Inverse square root and Newton-Raphson iteration}

We use the same NR-iteration as in the previous section. Since we do
not need the data conversion between single and double precision
formats, the code here actually becomes simpler. List  \ref{list:mixedNR}
gives this part of the code.

\begin{lstlisting}[caption={Calculation of inverse square root in {\tt
v4sf}.}, label={list:mixedNR}]
v4sf r2, rinv;

rinv = __builtin_ia32_rsqrtps(r2);
r2 *= rinv; /* start Newton-Raphson */
r2 *= rinv;
V4SF_OP_SCALAR(r2, -= , 3.0f);
rinv *= r2; /* now, rinv = -2.0/r */
\end{lstlisting}

\subsection{Accumulating acceleration and potential}

Before accumulation of acceleration and potential,
we now have to convert single-precision data back
to double precision. Before doing so, we need to split one piece of 128-bit
data with 4 single-precision words into two (effectively) 64-bit words
with two single-precision words. This is done by the {\tt movhlps}
instruction. Then we convert the result to double precision using a {\tt
cvtps2pd} instruction. The code appears in list \ref{list:accum}.

\begin{lstlisting}[caption={Accumulating x element of acceleration.}, label={list:accum}]
v4sf x, rinv3, ftmp;
v2df ax;

x *= rinv3;
ftmp = __builtin_ia32_movhlps(ftmp, x); 
ax += __builtin_ia32_cvtps2pd(x); 
ax += __builtin_ia32_cvtps2pd(ftmp);
\end{lstlisting}

For the jerk, we directly accumulate them in quadruples of 
single-precision words, since we do not need double-precision accuracy for jerk.
After total force is obtained, we add up higher two words and lower two words
of the accumulated data.

\subsection{The whole code}\label{subsec:wholecode}

List \ref{list:mixed}  
shows the entire code. 
We provide a simple library which one can call to use this function
in a way similar to the way the GRAPE hardware is used.
We show an example $N$-body code using this library in Appendix
\ref{app:nbody}. 
Note that actual code in list \ref{list:mixed} is slightly different from
code in list (\ref{list:mixedjp}-\ref{list:accum}).
We use arrays instead of vector types in the $j$ particle structure. 
Codes in list \ref{list:subpd} and \ref{list:accum} are abstracted
into macros for readability and saving of registers.

\begin{lstlisting} [caption={
	Total code of mixed precision force calculation.
}, label={list:mixed}]
static v4sf frv_coeff = {0.75, 0.75, 0.75, 0.75};
static v2df PotCorrect = {-0.5, -0.5};
static v2df AccCorrect = {-0.125, -0.125};
static v4sf V4sfEps2  = {1./65536., 1./65536., 1./65536., 1./65536.};
static v2df EpsInv = {256.0, 256.0};
static int Nbody;
static pJppack Jptr;

void SSEGRAV_Initialize(int nbody, double eps2){
    double bias = rsqrt_bias(frsqrt_1NR); /* measure bias */
    double epsinv = 1.0/sqrt(eps2);

    V2DF_OP_SCALAR(PotCorrect, =, -0.5*(1.0-bias));
    V2DF_OP_SCALAR(AccCorrect, =, -0.125*(1.0 - 3.0*bias));
    Eps2 = eps2;
    V2DF_OP_SCALAR(V2dfEps2, =, eps2);
    V4SF_OP_SCALAR(V4sfEps2, = , (float)eps2);
    V2DF_OP_SCALAR(EpsInv, =, epsinv);
    V4SF_OP_SCALAR(frv_coeff, =, 0.75*(1.0 - 2*bias));
    Nbody = nbody;
    Jptr = memalign(64, ((1+nbody)/2) * sizeof(Jppack));
}

#define SUB_PACK(fx, ptr, xi0, xi1, ftmp){ \
	fx    = __builtin_ia32_cvtpd2ps(*(v2df*)(ptr) - xi0); \
	ftmp  = __builtin_ia32_cvtpd2ps(*(v2df*)(ptr) - xi1); \
	fx    = __builtin_ia32_unpcklps(fx, ftmp); \
}

#define ACCUMLATE(phi, op, fx, ftmp){ \
	ftmp  = __builtin_ia32_movhlps(ftmp, fx); \
	phi op  __builtin_ia32_cvtps2pd(fx); \
	phi op  __builtin_ia32_cvtps2pd(ftmp); \
}

static inline v2df hadd_ps2pd(v4sf x){
	v4sf tmp = __builtin_ia32_movhlps(tmp, x);
	return __builtin_ia32_cvtps2pd(x) + __builtin_ia32_cvtps2pd(tmp);
}

void SSEGRAV_CalcAccJerkPot(pAccJerk ajout, int *index){
	int j, k;
	int nj = Nbody;
	v2df ax, ay, az, pot;
	v4sf jx, jy, jz;
	pJppack pr  = Jptr, prj = Jptr;
	int i0 = index[0], i1 = index[1];
	static v2df zero = {0.0, 0.0};
	static v4sf fzero = {0.0, 0.0, 0.0, 0.0};
	v2df xi0[3], xi1[3];
	v4sf vi[3];
	v4sf feps2 = V4sfEps2;
	
	pot = ax = ay = az = zero;
	jx = jy = jz = fzero;

	/* set i-particle */
	for(k=0;k<3;k++){
		float vi0 = pr[i0/2].v[k][2*(i0%2)];
		float vi1 = pr[i1/2].v[k][2*(i1%2)];
		
		V2DF_OP_SCALAR(xi0[k], =, pr[i0/2].x[k][(i0%2)]);
		V2DF_OP_SCALAR(xi1[k], =, pr[i1/2].x[k][(i1%2)]);
		V4SF_OP_VECTOR(vi[k],  =, vi0, vi1, vi0, vi1);
	}

	/* force loop */
	for(j=0;j<nj;j+=2,prj++){
		v4sf x,y,z,vx,vy,vz;
		v4sf r2, rv, rinv, rinv2, rinv3;

		__builtin_prefetch(prj+2, 0, 3);
		__builtin_prefetch(8+(double *)(prj+2), 0, 3);

		SUB_PACK(x, prj->x[0], xi0[0], xi1[0], vx);
		SUB_PACK(y, prj->x[1], xi0[1], xi1[1], vy);
		SUB_PACK(z, prj->x[2], xi0[2], xi1[2], vz);

		vx = *(v4sf*)(prj->v[0]) - vi[0];
		vy = *(v4sf*)(prj->v[1]) - vi[1];
		vz = *(v4sf*)(prj->v[2]) - vi[2];

		r2 = x*x + y*y + z*z + feps2;
		rv = vx*x + vy*y + vz*z;

		rinv = __builtin_ia32_rsqrtps(r2);
		r2 *= rinv;
		r2 *= rinv;
		V4SF_OP_SCALAR(r2, -= , 3.0);
		rinv *= r2;

		rinv2 = rinv * rinv;
		rv *= frv_coeff; /* 3/4(1-2bias) */
		rv *= rinv2;
		rinv *= *(v4sf *)(prj->m);

		rinv3 = rinv * rinv2;

		vx *= rinv3; jx += vx;
		vy *= rinv3; jy += vy;
		vz *= rinv3; jz += vz;
		
		ACCUMLATE(pot, -=, rinv, vx);
		x *= rinv3;
		y *= rinv3;
		z *= rinv3;
		ACCUMLATE(ax, +=, x, vx);
		ACCUMLATE(ay, +=, y, vy);
		ACCUMLATE(az, +=, z, vz);

		x  *= rv; jx -= x;
		y  *= rv; jy -= y;
		z  *= rv; jz -= z;
	}

	/* post loop procedure */
	pot *= PotCorrect; /* -1/2(1-bias) */
	{
		v2df mass = {pr[i0/2].m[2*(i0%2)], pr[i1/2].m[2*(i1%2)]};
		pot += mass * EpsInv; /* phi_i += m_i/eps */
	}
	ax *= AccCorrect; /* -1/8(1-3bias) */
	ay *= AccCorrect;
	az *= AccCorrect;
	{
		v2df djx, djy, djz;
		djx = hadd_ps2pd(jx) * AccCorrect;
		djy = hadd_ps2pd(jy) * AccCorrect;
		djz = hadd_ps2pd(jz) * AccCorrect;

		V2DF_SCATTER(ax, ajout[0].a,   ajout[1].a);
		V2DF_SCATTER(ay, ajout[0].a+1, ajout[1].a+1);
		V2DF_SCATTER(az, ajout[0].a+2, ajout[1].a+2);
		V2DF_SCATTER(djx, ajout[0].j,   ajout[1].j);
		V2DF_SCATTER(djy, ajout[0].j+1, ajout[1].j+1);
		V2DF_SCATTER(djz, ajout[0].j+2, ajout[1].j+2);
		V2DF_SCATTER(pot, &ajout[0].pot, &ajout[1].pot);
	}
}
\end{lstlisting}

\subsection{Performance}

\begin{table}
\caption{
	Performance of list \ref{list:mixed} in cycles-per-interaction
	and Gflops, on Athlon 64 3000+ 2.0GHz, when $N=1024$.
	The first column is performance of the output by GCC 3.3.1
	with option -O3, the second is that of hand-tuned assembly code
	after GCC.
}
\label{table:mixedperf}
\begin{tabular}{r r}

\hline
\multicolumn{1}{l}{GCC} & \multicolumn{1}{l}{Hand-tuning} \\
\hline
 30.6 cycles & 29.6 cycles\\
 3.92 Gflops & 4.05 Gflops \\ 
\hline
\end{tabular}
\end{table}

Table \ref{table:mixedperf}  
gives the performance of the
code listed above.  The second column gives the performance of the
code after hand-tuning the assembly output.

The best performance we have measured is 4.05 Gflops, or 3.19 times faster
than the original C code described in section 2.

\section{Discussion and Summary}

In this paper we describe in detail various ways of improving the
performance of the force-calculation loop for gravitational
interactions between particles. Since modern microprocessors have many
instructions which cannot be easily exploited by existing compilers,
we can achieve quite a significant performance improvement if we write
a few small library in assembly language and/or using instruction-set specific
extensions for the compiler offered to us.

Our implementation will give a significant speedup for almost any
$N$-body integration program.  In addition, we believe that similar
optimization is possible in many other compute-intensive applications,
within astrophysics as well as in other areas of physics and science
in general.

The source code and documentation are available at:\\
{\tt http://grape.astron.s.u-tokyo.ac.jp/\~{}nitadori/phantom/}
\section{Acknowledgment}
We thank Jumpei Niwa for making his Opteron machines available for
the development of the optimized force loop and also for his helpful
advice.
We are also grateful to Eiichiro Kokubo and Toshiyuki Fukushige
for their encouragement during the development of the code.
We would like to thank all of those who have been involved
in the GRAPE project,
which has given us a lot of hints for speeding-up of the gravity calculation.
P.H. thanks Prof. Ninomiya for his kind hospitality at the Yukawa
Institute at Kyoto University, through the Grants-in-Aid for
Scientific Research on Priority Areas, number 763, "Dynamics of
Strings and Fields", from the Ministry of Education, Culture, Sports,
Science and Technology, Japan. 
This research is partially supported by the Special Coordination Fund
for Promoting Science and Technology (GRAPE-DR project), also from the
Ministry of Education, Culture, Sports, Science and Technology, Japan.

\section{Sample $N$-body code using Hermite hierarchical time step scheme}
\label{app:nbody}
We show sample $N$-body code using Hermite hierarchical time step scheme
and the library in subsection \ref{subsec:wholecode}.

\begin{lstlisting}
#include <stdio.h>
#include <stdlib.h>
#include <malloc.h>

#define PREFETCH(addr)  __builtin_prefetch(addr, 0, 3);
#define PREFETCHW(addr) __builtin_prefetch(addr, 1, 3);
#define MEGAHELTZ 2000

struct AccJerk{
	double a[3];
	double j[3];
	double pot;
};

struct Particle{
	double x[3];
	double v[3];
	double a[3];
	double j[3];
	double m;
	double pot;
	double time;
	double timestep;
}; /* 128-byte */

void CalcAccJerkUsing_iIndex
(int nbody, int *iIndex, int iNum, struct AccJerk *ajout) {
	int NumPipe = 2;
	int iNumRest = iNum;

	while(iNumRest){
		int ni = MIN(NumPipe, iNumRest);
		if(ni == 1) iIndex[1] = iIndex[0];

		SSEGRAV_CalcAccJerkPot(ajout, iIndex, nbody);

		iNumRest -= ni;
		iIndex += ni;
		ajout += ni;
	}
}

int GravityDriver(char *parmfile){
	int i, k;
	int nbody;
	double epsinv, TimeEnd, dEOutTime, MaxStep, Eta_s, SqrtEta;
	double SystemTime = 0.0;
	double eps2;
	double InitEnergy;
	FILE *fpin = NULL, *fpout = NULL;
	struct Particle *ptcl;
	struct AccJerk *ajnew;
	int *iIndex;
	long NumPtclStep = 0, NumStep = 0;
	long tsc0, tsc1;
	double cycle;
	long CyclePredict = 0, CycleForce = 0
         CycleCorrect = 0, CycleTot = 0;
	double WallTime;

	GetParam(parmfile, &epsinv, &TimeEnd, &MaxStep,
			 &dEOutTime, &Eta_s, &SqrtEta, &fpin, &fpout);
	if(epsinv == 0.) eps2 = 0.;
	else eps2 = 1.0/(epsinv*epsinv);

	assert(fpin != NULL);
	assert(1 == fscanf(fpin,"%d\n", &nbody));
	printf("N = %d\n", nbody);

	ptcl = memalign(64, nbody * sizeof(struct Particle));
	ajnew = memalign(64, (1+nbody) * sizeof(struct AccJerk));
	iIndex  = calloc(1+nbody , sizeof(int));

	ReadSnapshot(fpin, nbody, ptcl, &SystemTime);

	SSEGRAV_Initialize(nbody, eps2/*, epsinv*/);
	BSTEP_Initialize(nbody, 0.0, MaxStep);

	SSEGRAV_Predict(ptcl, SystemTime/*, nbody*/);

	/* initial step */
	for(i=0;i<nbody;i++) iIndex[i] = i;
	rdtscll(tsc0);
	CalcAccJerkUsing_iIndex(nbody, iIndex, nbody, ajnew);
	rdtscll(tsc1);
	cycle = (tsc1 - tsc0) / ((double)nbody * nbody);
	printf("%f cycle per interruction, %F Gflops\n\n",
		    cycle, MEGAHELTZ/1.e3/cycle*60.0);

	for(i=0;i<nbody;i++){
		for(k=0;k<3;k++){
			ptcl[i].a[k] = ajnew[i].a[k];
			ptcl[i].j[k] = ajnew[i].j[k];
		}
		ptcl[i].pot = ajnew[i].pot;
		InitTimestep(ptcl+i, Eta_s, MaxStep);
		BSTEP_PushAParticle(i, ptcl[i].timestep, &ptcl[i].timestep);
	}
 	InitEnergy = GetEnergy(ptcl, nbody);
	PrintEnegy(ptcl, nbody, InitEnergy, SystemTime);

	/* evolve */
	rdtscll(tsc0);
	CycleTot -= tsc0;
	CycleCorrect -= tsc0;
	while(1){
		int iNum;

		BSTEP_PullParticles(iIndex, &iNum, &SystemTime);
		rdtscll(tsc0);
		CycleCorrect += tsc0;
		if(SystemTime >= TimeEnd) break;

		CyclePredict -= tsc0;
		SSEGRAV_Predict(ptcl, SystemTime/*, nbody*/);
		rdtscll(tsc0);
		CyclePredict += tsc0;

		CycleForce -= tsc0;
		CalcAccJerkUsing_iIndex(nbody, iIndex, iNum, ajnew);
		rdtscll(tsc0);
		CycleForce += tsc0;

		CycleCorrect -= tsc0;
		for(k=0;k<iNum;k++){
			int ii = iIndex[k];
			char *addr = (char *)(ptcl + iIndex[k+1]);
			PREFETCHW(addr);
			PREFETCHW(addr+64);
			PREFETCH(ajnew+k+1);
			HermiteCorrect(ptcl+ii, ajnew+k, SqrtEta);
			BSTEP_PushAParticle
				(ii, ptcl[ii].timestep, &ptcl[ii].timestep);
		}
		if(fmod(SystemTime, dEOutTime) == 0.0){
			for(i=0;i<nbody;i++) ptcl[i].pot = ajnew[i].pot;
			PrintEnegy(ptcl, nbody, InitEnergy, SystemTime);
		}
		NumPtclStep += iNum;
		NumStep++;
	}
	rdtscll(tsc1);
	CycleTot += tsc1;

	WallTime = (double)CycleTot/MEGAHELTZ / 1.E6;
	printf("Wall clock time: %f sec\n", WallTime);
	printf("predict: %f usec/blockstep, %f cycle/loop\n",
			(double)CyclePredict/NumStep/MEGAHELTZ,
			(double)CyclePredict/(NumStep*nbody));
	printf("force  : %f usec/blockstep, %f cycle/loop\n",
			(double)CycleForce/NumStep/MEGAHELTZ,
			(double)CycleForce/(NumPtclStep*nbody));
	printf("correct: %f usec/blockstep, %f cycle/loop\n",
			(double)CycleCorrect/NumStep/MEGAHELTZ,
			(double)CycleCorrect/(NumPtclStep));
	printf("%f particles/blockstep\n\n", (double)NumPtclStep/NumStep);

	/* final synchronizing step */
	SSEGRAV_Predict(ptcl, TimeEnd/*, nbody*/);
	for(i=0;i<nbody;i++) iIndex[i] = i;
	CalcAccJerkUsing_iIndex(nbody, iIndex, nbody, ajnew);
	for(i=0;i<nbody;i++){
		ptcl[i].pot = ajnew[i].pot;
		ptcl[i].timestep = TimeEnd - ptcl[i].time;
		HermiteCorrect(ptcl+i, ajnew+i, SqrtEta);
	}
	PrintEnegy(ptcl, nbody, InitEnergy, TimeEnd);

	if(fpout) WriteSnapshot(fpout, nbody, ptcl, TimeEnd);

	SSEGRAV_Finalize();
	BSTEP_Finalize();
	free(ptcl);
	free(ajnew);
	free(iIndex);
	
	fclose(fpin);
	fclose(fpout);
	return 0;
}

int main(){
	GravityDriver("initparm");
	return 0;
}
\end{lstlisting}


\end{document}